\documentclass[aps,pra,amsmath,amssymb,superscriptaddress,notitlepage]{revtex4-1}
\pdfoutput=1

\usepackage{amsmath}
\usepackage{amssymb}
\usepackage{amsthm}
\usepackage[pdftex]{color}
\usepackage{graphicx}% Include figure files
\usepackage{dcolumn} % Align table columns on decimal point
\usepackage{bm} % bold math
\usepackage{epic}
\usepackage{bbm}
\usepackage{hyperref}
\usepackage{longtable}
\usepackage{slashed}
\usepackage{times}
\usepackage{ulem}   % to strike things out
\begin{document}
\bibliographystyle{apsrev}

%\draft

\title{Response of fermions in Chern bands to spatially local quenches}

\author{Adolfo G. Grushin} 

\author{Sthitadhi Roy} 

\author{Masudul Haque} 

\affiliation{Max-Planck-Institut f\"ur Physik komplexer Systeme, N\"othnitzer Stra{\ss}e 38, 01187 Dresden, Germany}   

\begin{abstract}

  We study the dynamical evolution of Chern-band systems after subjecting them to local quenches.
  For open-boundary systems, we show for half-filling that the chiral nature of edge states is
  manifested in the time-dependent chiral response to local density quenches on the edge.  In the presence
  of power-law traps, we show how to mimic the half-filling situation by choosing the appropriate
  number of fermions depending on the trap size, and explore chiral responses of edges to local
  quenches in such a configuration.  We find that perturbations resulting from the quenches
  propagate at smaller group velocities as the gap controlling the spatial
extent of the edge modes decreases.  Our results provide
  different routes to check dynamically the non-trivial nature of Chern bands.

\end{abstract}

\maketitle

\section{Introduction}

Chern insulators are topological states of matter that host an intrinsic quantized quantum Hall
effect without the need of external magnetic fields.  Embedded today in the large body of
knowledge gathered around topological phases~\cite{HK10,QZ11}, Chern insulators were conceived soon
after the discovery of the quantum Hall effect \cite{H88} and can hold the key to realizing
interesting unconventional phenomena such as fractional many-body states in the absence of external
magnetic fields~\cite{BL13,PRS13} or more exotic electrodynamical phenomena such as a repulsive
Casimir effect~\cite{RG14}.

The equilibrium signatures of Chern insulators are by now well established. 
Their anomalous transport is determined by a topological invariant associated to each band 
known as the Chern number related to the Hall conductivity through~\cite{NSO10}
$\sigma_{xy}=\sum_{a}C_{a}e^2/h$, where
$C_a$ is the Chern number for each filled band $a$. The non-trivial band topology, i.e. non-zero
Chern numbers, implies the existence of propagating chiral modes localized at the edge of a finite
sample.
The recent experimental realization of solid-state materials~\cite{CZJ13,KGF14,CYT14,BFK15,CZK15} and optical-lattice
systems~\cite{BDZ08,AAL13,UJM13,MSK13,JMD14,ALS14} having Chern bands (or closely related
properties) has boosted the interest in such systems. 

Compared to equilibrium behaviors, out-of-equilibrium features of Chern bands have attracted
interest only recently, particularly because realizations with ultracold atoms are well-suited for
studying real-time dynamics.  It is therefore experimentally relevant to characterize topological
states of matter not only with transport signatures but also with dynamical
measurements~\cite{PC12,GDD13,DG13,PC13,HLE14,ALS14}.  For instance, the Chern number can be
determined dynamically by following the trajectory of a wave packet in a tilted optical
lattice,~\cite{PC12,DG13} a method which was successfully used experimentally recently.
\cite{JMD14,ALS14} It was also shown theoretically that edge state dynamics can be visualized by
tailoring external potentials that, when released, provide direct imaging of the chirality of the
Chern bands through the chiral propagation of the atoms in a hard-wall trap~\cite{GDD13}.  Global
quenches in Chern band models and related systems have been addressed in a number of recent
studies~\cite{KP12,KTP12,PSD13,HLE14,CCB15,DR14,S14}.

%%%%%%%%%%% FIGURE %%%%%%%%%%%%%%%%% FIGURE  %%%%%%%%%%%%%%%%%% FIGURE %%%%%%%%  
\begin{figure}
   \centering
   \includegraphics[width=0.5\columnwidth]{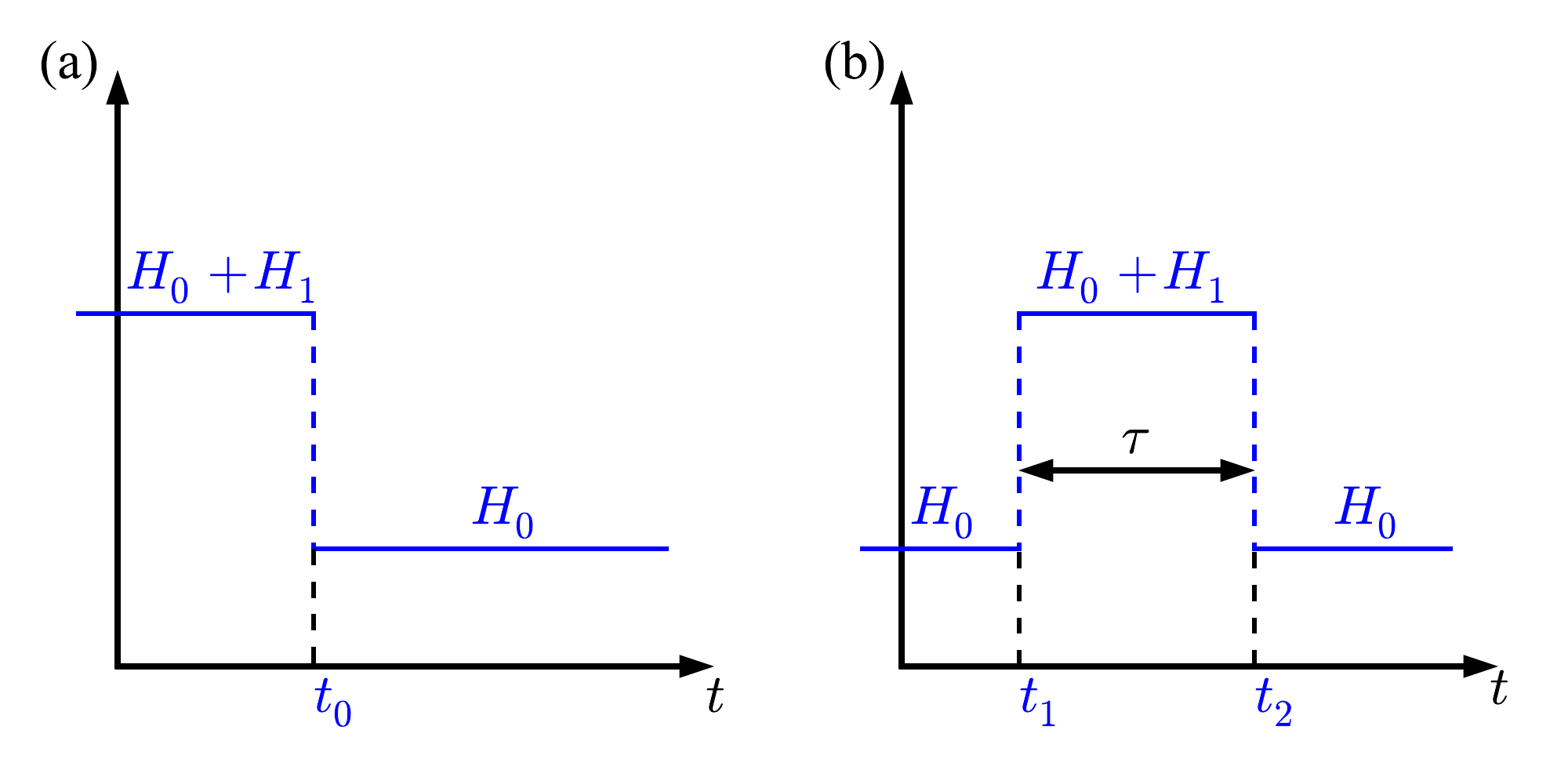}
   % requires the graphicx package
   \caption{(color online) The figures schematically show the two quench protocols 
   %The two schematic time dependent perturbations
     considered in this work: (a) is a quench from an equilibrium state at $t=t_0$ while (b) is a pulse of width
     $\tau=t_{2}-t_{1}$.}
   \label{fig:quenches}
\end{figure}
%%%%%%%%%%% FIGURE %%%%%%%%%%%%%%%%% FIGURE  %%%%%%%%%%%%%%%%%% FIGURE %%%%%%%%  

In this work, we complement this body of knowledge by studying the effect of \emph{local} quenches
on Chern bands filled with fermions.  We focus particularly on the real-time chiral response of the
edges.  This study thus supplements existing proposals for experimental protocols to probe the
chiral nature of such states~\cite{GSN10,LLC10,SGD10,KP12,KTP12,BCH12,LLN13,HLE14,DSD14,GBG12}. 

We consider two kinds of quantum quenches shown schematically in Fig.~\ref{fig:quenches}: a single
quench and a pulse consisting of two sudden changes of the Hamiltonian.  The background Hamiltonian
$H_{0}$ is taken to be a tight-binding model with topologically non-trivial bands.  The perturbing
Hamiltonian $H_{1}$ is chosen to be a localized density perturbation of the system.  We will
concentrate on perturbations at one site at the edge of the fermionic system, in order to
dynamically probe the chirality of the edge states.  In the first case, we start with the ground
state of $H_0+H_1$, and then remove the local perturbation $H_1$ and follow time evolution under the
Hamiltonian $H_0$.  In the second case we start with the ground state of $H_0$ and turn on the local
perturbation $H_1$ for a finite time interval $\tau$.

We first consider (section \ref{sec:half-filling}) half-filling of an open-boundary system.  In this
case, the edge of the fermionic system is simply the edge of the lattice.  Next, motivated by
experiments with cold atomic systems, we explore the effect of power-law traps (section
\ref{sec:trap}), i.e., traps of the form $\sim r^{\gamma}$.  In cold-atom experiments, the atoms are
usually loaded in a harmonic trap, i.e, $\gamma=2$.  Since systems in harmonic traps often do not
have a well-defined sharp edge, there is significant interest in power-law traps with large
exponents $\gamma$, which is expected to be more similar to a system with sharp
boundaries~\cite{GDD13,DG13,GSG13,MSH05}.  Accordingly, we consider fermions in a Chern band in the
presence of a power-law trapping potential $J(r/r_0)^{\gamma}$, where $J$ is the hopping energy that
sets the energy width of the bands.  The parameter $r_0$ then functions as a trapping length scale,
determining the spatial extent of the trapped fermionic cloud~\cite{GDD13,DG13,ALS14}.  
We first characterize
the equilibrium properties in such a trap and then study the effect of local quenches on it.

Our main findings are as follows.  When the bands are topologically non-trivial, a local quench of a
site potential at the edge generates a spatially localized pulse that propagates chirally around the
edge of the sample.  This is a straightforward and very direct manifestation of the topological
nature of $H_{0}$.  The front of the pulse has speed determined by the hopping scale $J$, but the
peak of the pulse has smaller speed for smaller bulk gaps at the relevant points in the Brillouin Zone.  We attribute this phenomena to the
increasingly poor localization of the edge states as the gap decreases; the perturbation is spread
out over increasingly more lattice sites rendering the sharp boundary picture less accurate.  In the
second part, considering power-law traps with large exponents $\gamma$, we show how the trap
parameter $r_0$ determines an optimal particle number for mimicking the half-filling situation, and
hence for observing the chiral nature of the edge states.

The paper is structured as follows.  In section \ref{sec:model} we present a generic model defined
on the square lattice that has both trivial insulator and Chern insulator phases; we use this
model throughout the paper.  Next, in section \ref{sec:half-filling} we consider a half-filled
rectangular lattice with open boundaries and study local perturbations at the edge, of the two forms
shown in Fig.~\ref{fig:quenches} (quench and pulse).  In section \ref{sec:trap} we consider
power-law potential traps.  We exemplify the decisive role of $r_{0}$ in selecting the optimal
fermion number in order to mimic a half-filled system with a populated edge.  We show how the
non-equilibrium features (chiral propagation) survive in this case.  In section \ref{sec:conc}, we
summarize and provide context, and discuss the experimental set-ups that could access the chiral
nature of the Chern bands as proposed here.
Since the Hamiltonians are non-interacting, the methods for finding ground states and time-evolving
many-body states are standard, and are briefly described in Appendix~\ref{app:evol}.

\section{\label{sec:model}Model Hamiltonian and quench protocols}

%%%%%%%%%%% FIGURE %%%%%%%%%%%%%%%%% FIGURE  %%%%%%%%%%%%%%%%%% FIGURE %%%%%%%%  
\begin{figure}
   \centering
   \includegraphics[width=0.5\columnwidth]{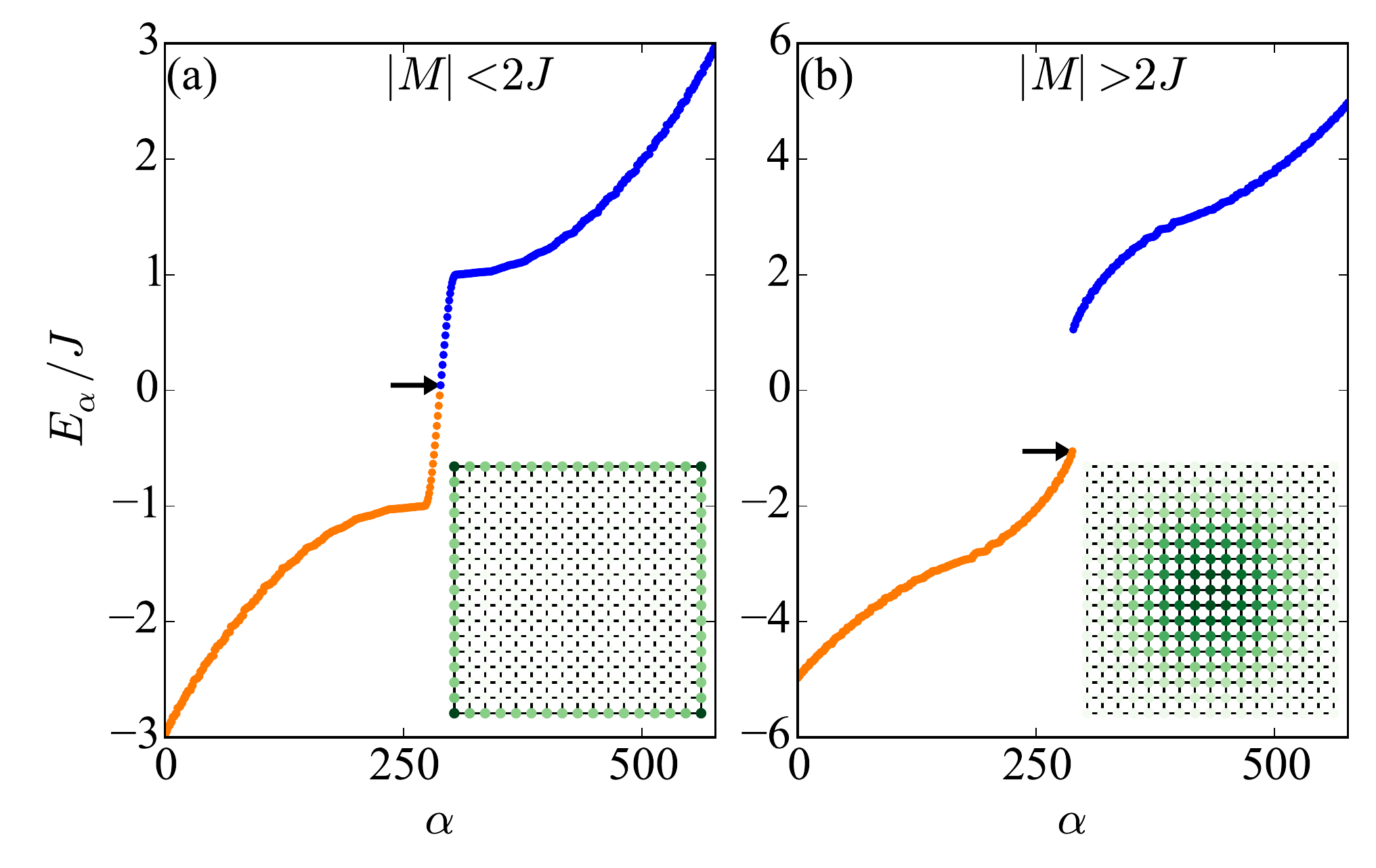} 
   \caption{(color online) The single-particle energy spectrum for the model \eqref{eq:CI} on a
     $17\times17$ lattice (each site has two orbitals) with open boundary conditions.  The
     parameters are chosen to be in (a) the Chern insulator phase and (b) the trivial insulator
     phase. The insets show the real space probability distribution of the single-particle
     eigenstates closest to half-filling (marked by the arrows on the spectrum) with the color
     intensity showing the probability at any particular site.}
   \label{fig:bands}
\end{figure}
%%%%%%%%%%% FIGURE %%%%%%%%%%%%%%%%% FIGURE  %%%%%%%%%%%%%%%%%% FIGURE %%%%%%%%  

We are interested in the dynamics of a generic two band Chern insulator model in two dimensions.  As
a representative of a Chern-band system, we will consider the following single particle Hamiltonian
defined on the square lattice~\cite{BHZ06,QHZ08} with a spin-1/2 degree of freedom (either orbital or
spin)
\begin{subequations}
\label{eq:CI}
\begin{eqnarray}
\nonumber
H_{\mathrm{CI}}&=&H_{J}+H_{M},\\
H_{J}&=&-J\sum_{i}\big[c^{\dagger}_{i}\dfrac{(\sigma_{z}-i\sigma_{x})}{2}c_{i+\hat{x}}+c^{\dagger}_{i}\dfrac{(\sigma_{z}-i\sigma_{y})}{2}c_{i+\hat{y}}+h.c.\big],\\
H_{M}&=&Mc^{\dagger}_{i}\sigma_{z}c_{i}, 
\end{eqnarray}
\end{subequations}
with hopping $J$ between the sub-lattices and a staggered chemical potential $M$.  The fermionic
operators $c_i$, $c^{\dagger}_i$ for each site $i$ represents a two-spinor of the operators for each
orbital/spin on that site, e.g., $c^{\dagger}_i=\left( c^{\dagger}_{i\uparrow}, c^{\dagger}_{i\downarrow}\right)$.  Summations over the spin index are thus implicit.

In Fourier space, a  two band model Hamiltonian can be written generally as 
\begin{equation}
\label{eq:CIkspace}
H_{\mathrm{CI}}(\mathbf{k})=\epsilon_{\mathbf{k}}+\boldsymbol{\sigma}\cdot\mathbf{d}_{\mathbf{k}}. 
\end{equation}
In our case, $\mathbf{d_k}$ and $\epsilon_{\mathbf{k}}$ are 
\begin{subequations}
\begin{eqnarray}
\label{eq:CIkinterm1}
d_{x,\mathbf{k}}&=& -J\sin(k_{x}),\\
\label{eq:CIkinterm2}
d_{y,\mathbf{k}}&=& -J\sin(k_{y}),\\
\label{eq:CImassterm}
d_{z,\mathbf{k}}&=&M-J\sum_{i=x,y}\cos(k_{i}),~\text{and}~ \epsilon_{\mathbf{k}}= 0.
\end{eqnarray}
\end{subequations}
At half-filling such a single particle Hamiltonian represents a Chern insulator with chiral edge
states whenever $-2J<M<2J$.  The topological and trivial eigenstates for a finite sample are shown
in Fig.~\ref{fig:bands} (a) and (b) respectively, together with the real space probability
distribution of the eigenstates closest to zero energy ($E=0$). In the trivial case, there is no
sign of edge states whereas for the Chern insulator, the edge states are clearly visible.  At
half-filling the Chern insulator state has $\sigma_{xy}=Ce^2/h$ where $C=\pm1$ is the Chern number
of the filled lower band.  We note that $\epsilon_{\mathbf{k}}=0$ does not affect the topological
nature of the state and thus it is irrelevant for our discussion.

The Hamiltonian \eqref{eq:CI} (with small modifications) has been used widely as a basic standard
model for two-band systems with each model having a Chern number~\cite{BHZ06,QHZ08,WBR12}. The
Haldane honeycomb model \cite{H88} can be cast into this form if the two sites in the honeycomb unit
cell are mapped into the two orbitals on the same site of Hamiltonian \eqref{eq:CI}.  The results we
present are quite generic, and should hold qualitatively for other Chern insulator models such as
the Haldane model~\cite{H88}, and also for other systems having bands with nonzero Chern number,
such as the Hofstadter model~\cite{H76}.

%% which have experimentally been achieved in bosonic cold-atomic lattices with externally
%% driving~\cite{JMD14,AAL13,ALS14}.

In section \ref{sec:half-filling} we will set $H_{0}=H_{\mathrm{CI}}$ while in section \ref{sec:trap} a trapping potential $H_{\mathrm{trap}}$ (to be defined later on) 
will be added to the Hamiltonian: $H_{0}=H_{\mathrm{CI}}+H_{\mathrm{trap}}$.

We will be concerned with physical processes where the system described by Hamiltonian $H_{0}$ is
acted upon by two types of perturbation.  The first is of the form
\begin{equation}
\label{eq:quench}
H_{\mathrm{e}}(t) = H_{1}\theta(t_{0}-t),
\end{equation}
which we will refer to as a \emph{quench from equilibrium} [see Fig.~\ref{fig:quenches}(a)]
and the second is defined as 
\begin{equation}
\label{eq:pulse}
H_{\mathrm{e}}(t) = H_{1}\left[\theta(t-t_{1})-\theta(t-t_{2})\right],
\end{equation}
that is non-zero for a time $\tau=t_{2}-t_{1}$,
and we refer to as a \emph{pulse} [see Fig.~\ref{fig:quenches}(b)].
We focus on local density perturbations that we label $H_{1}$. We define a local
density perturbation as an increase or deficit of the charge density around a particular site $l$ with magnitude $\mu_{l}$
\begin{eqnarray}\label{eq:chargelike}
H_{1}^{l} &=& \mu_{l} c^{\dagger}_{l}c_{l},
\end{eqnarray}
Here no implicit summation is assumed over the site index, but there is an automatic summation over
the spin index, since $c_l$ and $c^{\dagger}_l$ are two-spinors. 

In the concluding section we will discuss briefly possible physical implementations of these types
of perturbations in cold-atomic experiments and in condensed matter settings.  

Next, we perform time evolution by evolving the one-particle density matrix, or the matrix of
correlators, $\varrho_{ij}=\langle{c^{\dagger}_i}{c_j}\rangle$ (see Appendix A for details); here
$i$, $j$ are site indices and the spin indices are implicit.  We present our results by plotting the
total densities at each site $i$, which correspond to the diagonal terms of the correlation matrix,
$\varrho_{ii}$, relative to $\varrho_{ii,0}$, the density at site $i$ calculated from $H_{0}$, with
spin summation implied.

%%%%%%%%%%%% FIGURE %%%%%%%%%%%%%%%%% FIGURE  %%%%%%%%%%%%%%%%%% FIGURE %%%%%%%%  
\begin{figure}
   \centering
   \includegraphics[width=0.5\columnwidth]{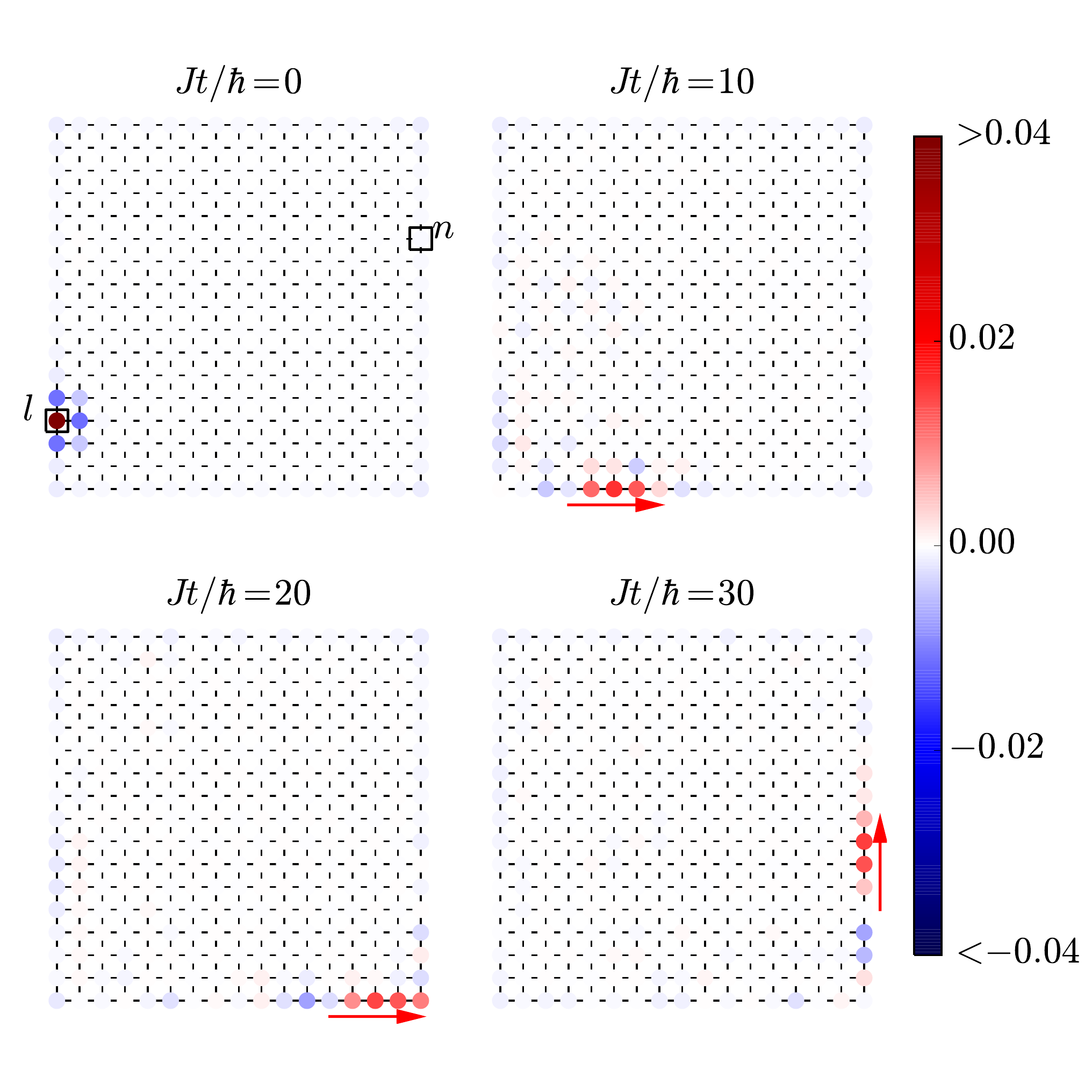} % requires the graphicx package
   \caption{(color online) The time evolution of the density in the Chern insulator, following a
     quench from equilibrium [Fig.~\ref{fig:quenches}(a)] with $t_0=0$.  Snapshots of the local
     local density difference $\varrho_{ii}(t)-\varrho_{ii,0}$ at each site $i$ are shown at four
     different times. A red (blue) color on any site denotes an excess (deficit) in fermion density
     relative to $\varrho_{ii,0}$ with the intensity of the color showing the magnitude of the
     difference. The arrows show the direction of the chiral propagation of the excitations. The
     local quench is performed at the site labeled by $l$ and the density measurements in
     Fig.~\ref{fig:edgepert2} are done at the site labeled $n$. For this simulation $\mu_l=-J/4$
     and $M/J=1$. Please see supplementary material for a video of the time evolution.}
   \label{fig:edgepert}
\end{figure}
%%%%%%%%%%%% FIGURE %%%%%%%%%%%%%%%%% FIGURE  %%%%%%%%%%%%%%%%%% FIGURE %%%%%%%%  

\section{\label{sec:half-filling} Half-filled open-boundary system}

In this section we consider rectangular open-boundary systems (without a trap) with $L=L_{x}\times L_{y}$ sites at half filling. We explore the chirality of the topological edge states through perturbations of the form \eqref{eq:chargelike}.

\subsection{Local density quench from equilibrium}

%%%%%%%%%%%% FIGURE %%%%%%%%%%%%%%%%% FIGURE  %%%%%%%%%%%%%%%%%% FIGURE %%%%%%%%  
\begin{figure}
   \centering
   \includegraphics[width=0.5\columnwidth]{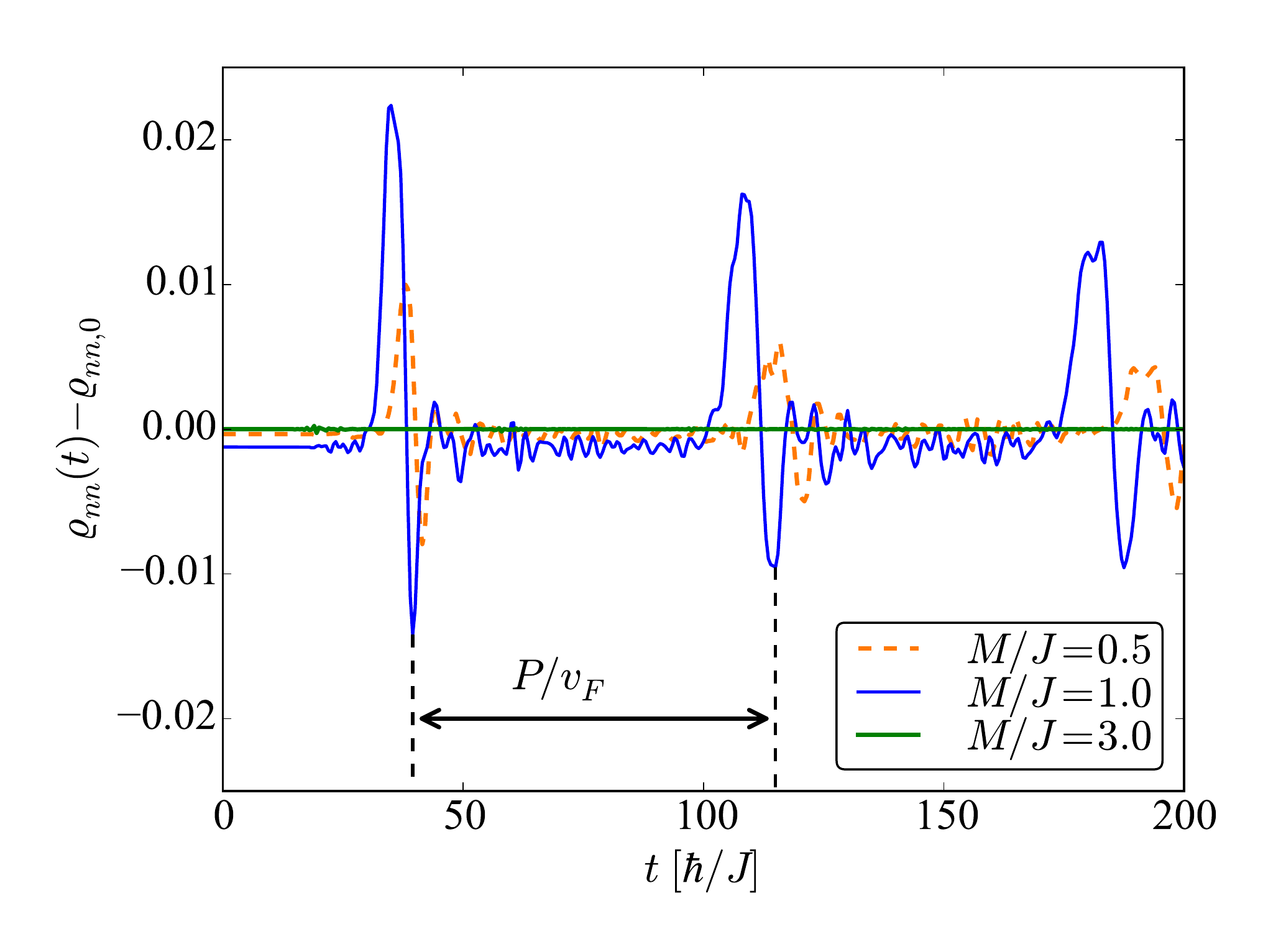} 
   \caption{(color online) The density difference $\varrho_{nn}(t)-\varrho_{nn,0}$ at the site $n$
     marked in Fig.~\ref{fig:edgepert} is plotted against time for a trivial insulator with $M/J=3$
     (featureless green solid line), a Chern insulator with $M/J=1$ (blue solid line) and a Chern
     insulator with $M/J=1/2$ (orange dashed line). The gap sizes at $\mathbf{k}=0$ are respectively
     $2|M-2J|=2J,2J$ and $J$, the local density perturbation parameter is $\mu_{l}=-J/4$.  Note the
     slight delay for the dashed curve, corresponding to a smaller gap at $\mathbf{k}=0$ and hence more extended
     edge state (see Eq.~\eqref{eq:loc}). The time interval $T_{\text{edge}} \simeq P/v_{F}$, where $P=64a$ is the perimeter of the lattice and $v_{F}=Ja/\hbar$ is the Fermi velocity of the edge state, is also shown by the black dashed lines.}
   \label{fig:edgepert2}
\end{figure}
%%%%%%%%%%%%% FIGURE %%%%%%%%%%%%%%%%% FIGURE  %%%%%%%%%%%%%%%%%% FIGURE %%%%%%%%  

We start with a local density quench, of the type of Fig.~\ref{fig:quenches}(a), at an edge site $l$,
\begin{equation}
H(t) = H_{0} + H_{1}^{l}\theta(t_{0}-t)  ,
\end{equation}
with $H_{0}=H_{\mathrm{CI}}$ and $H_{1}^{l}$ given by \eqref{eq:chargelike}.  Through the action of $H_{1}^l$ the initial ground
state has different density at site $l\in \mathrm{edge}$ compared to the ground state of $H_0$.  The
perturbation $\mu_{l}$ has equal effect on both spin components.  The particle number is
fixed to  $N=L_{x} \times L_{y}$, i.e. half-filling.   

In Fig.~\ref{fig:edgepert} we show the time evolution by presenting four snapshots of the local
local density difference $\varrho_{ii}(t)-\varrho_{ii,0}$ between the density at each site $i$
($\varrho_{ii}(t)$) and the one corresponding to the half-filled ground state of $H_0$
($\varrho_{ii,0}$).
It is clear from the snapshots that the perturbations travel along the edge of simulated lattice
with a well-defined chirality, in this case  counterclockwise.
The effect of the perturbation on the edge site labeled $n$, marked in Fig.~\ref{fig:edgepert}, is
shown in Fig.~\ref{fig:edgepert2}. The density difference at site $n$ shows a periodic pattern with
a period $T_{\text{edge}} \simeq P/v_{F}$, where $P$ is the perimeter of the lattice and $v_{F}$ is the Fermi
velocity of the edge state. The perimeter for this particular simulation is
$P=16a\times4=64a$.

%%%%%%%%%%%% FIGURE %%%%%%%%%%%%%%%%% FIGURE  %%%%%%%%%%%%%%%%%% FIGURE %%%%%%%%  
\begin{figure*}
   \centering
   \includegraphics[width=0.99\linewidth]{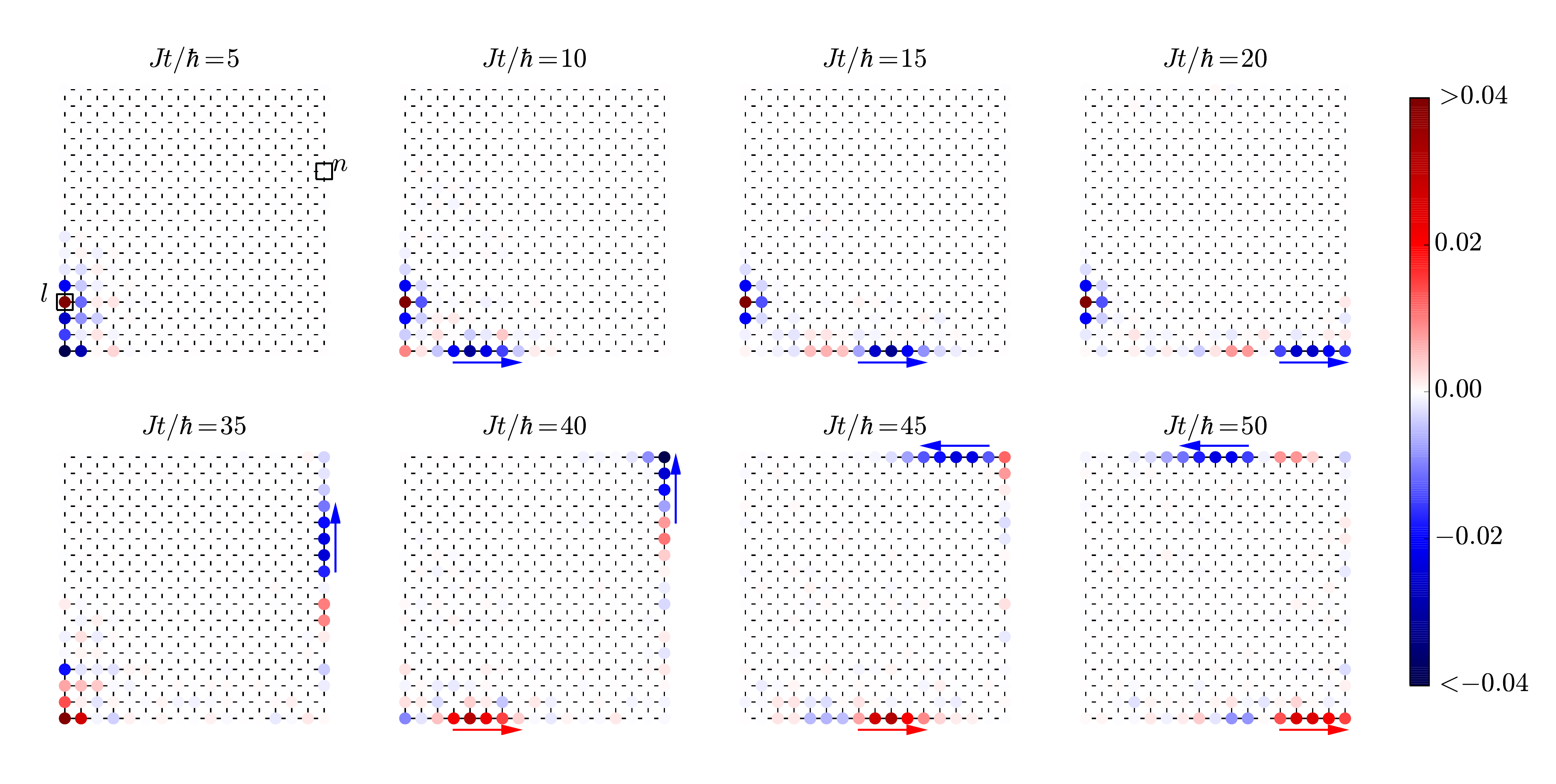} 
   \caption{(color online) The local density difference $\varrho_{ii}(t)-\varrho_{ii,0}$ is plotted
     for different times to show the dynamics of the edge excitation in the case of the pulse quench
     [Fig.\ref{fig:quenches}(b)] with $t_1=0$ and $t_2=30\hbar/J$. The red and blue arrows show the
     chiral propagation of the particle- and hole-like excitations respectively.  The quench
     protocol parameters are chosen such that $\tau>\hbar/J$.  Other parameters are the same as in
     Fig.~\ref{fig:edgepert}. Please see supplementary material for a video of the time evolution.}
   \label{fig:edgepertpulse1}
\end{figure*}
%%%%%%%%%%%% FIGURE %%%%%%%%%%%%%%%%% FIGURE  %%%%%%%%%%%%%%%%%% FIGURE %%%%%%%%  

In order to shed light on these numerical results, we estimate the value of $v_{F}$ and the localization length of the edge from the low energy theory of the model \eqref{eq:CI}. 
To do so, we first expand $H_0$ around each high-symmetry point of the Brillouin Zone, $ \mathbf{k}^{(i,j)}=\pi(i,j) $ with $i,j=0,1$.
The emergent effective low energy model is a massive Dirac equation around each one of them that takes the form
\begin{equation}
H^{(i,j)}(\mathbf{k})= -J \left[(-1)^{i}k_x+(-1)^{j}k_y\right]+m_{(i,j)}\sigma_{z},
\end{equation}
with four mass terms defined by
\begin{equation}
m_{(i,j)}= M-\left[(-1)^{i}+(-1)^{j}\right]J.
\end{equation}
If $|M/J|<2$, i.e. in the topological phase, only three out of the four mass terms have the same sign. On the other
hand when $|M/J|>2$, the trivial state, all of them have the same sign. 
For $M/J>0$ $(<0)$ the boundary between the Chern insulator and a trivial insulator (e.g. vacuum)
is modelled by choosing $m_{(0,0)}=m(y)$ ($m_{(\pi,\pi)}=m(y)$) such that $m(y)$ changes sign at the boundary, which we take it to be at $y=0$. The corresponding Dirac equation
has only $k_x$ as a good quantum number and has a gapless solution that decays exponentially as~\cite{LFS94}
\begin{equation}
\label{eq:loc}
\Psi_{\mathrm{edge}}(y) \sim e^{-\int^{y}_{0} m(y')dy'}\left[\begin{array}{c}1 \\1\end{array}\right].
\end{equation}
Fixing $M/J>0$, a sharp edge can be modeled by $m(y)=m_{(0,0)}\left[\theta(y)-\theta(-y)\right]$, that determines the
localization of the edge state to be inversely proportional to $|M-2J|$ (the case where $M/J<0$ is obtained by simply replacing $m_{(0,0)}\to m_{(\pi,\pi)}$).  Such an edge state, has a
dispersion $E= \pm v_{F}k_{y}$ ($E= \pm v_{F}k_{x}$) for edges along the $y$ ($x$) direction. The
sign is determined by the sign of the Chern number $C$ of the lower band and $v_{F}$ is set by the
bulk dispersion Fermi velocity. Therefore, $v_{F}$ can be read directly from \eqref{eq:CIkinterm1}.  For this model it is isotropic and takes the value $v_{F}=J$ (in units
of $\hbar$).

From this analysis it follows that, if $m_{(0,0)}$ is reduced, the edge states will have a finite
extent, having in general support on several rows close to the edge.  This in turn will affect the
period between the density pulses reaching a particular site, i.e., the period between the peaks in
Fig.~\ref{fig:edgepert2}.  Since the edge perturbation now has more sites to explore as it
propagates, one expects that the front of the propagating wave travels at the same speed as in a
narrow edge, but the peak of the density wave will travel more slowly due to the larger width of the
propagation channel.  This effect is shown in Fig.~\ref{fig:edgepert2} where the local density at a
site $n\in$ edge given by $\varrho_{nn}(t)$ is shown for two different instances within the Chern
insulator phase, corresponding to $M/J=1$ (blue-solid line) and $M/J=1/2$ (orange-dashed line).
Fig.~\ref{fig:edgepert2} shows how the highest crest of the oscillations shift to later times as
the $|M-2J|$ decreases from $J$ to $J/2$ as argued above.  The speed of the front of the wave is
apparently unaltered as expected because $v_F$ is independent of $m_{(0,0)}$.

Such an effect is observable in a simple toy model of the conducting chiral edge
by analyzing how a perturbation propagates along a conducting strip modeled as a
trivial tight binding square lattice of linear dimensions 
$l_{x}\times l_{y}$ with $l_{x}\gg l_{y}$, as a function of the strip thickness $l_{y}$. Although in this case the propagation is not chiral and 
thus an exact comparison is not possible, the wave packet indeed explores more sites as $l_{y}$ is increased, 
which effectively reduces the peak velocity in the $x$ direction, similar to what we observe for the Chern insulator edge.

The apparent decay of the main peak in Fig.~\ref{fig:edgepert2} on the other hand is likely to be due to
the spread of the wave packet along the edge rather than decay into the bulk. 
The latter is strongly suppressed by the initial perturbation being localized at a single edge site, having therefore 
no overlap with bulk states.

Fig.~\ref{fig:edgepert2} also shows data for the trivial insulator case, $M/J=3$ (featureless solid
green line).  There is no chiral propagation in this case due to the absence of edge states, so
there are no features observed in the density at site $n$.  

Note in addition that in Figs.~\ref{fig:edgepert} and ~\ref{fig:edgepert2} the perturbation satisfies $\mu_{l}<|M-2J|$.  We have
checked as well that our results hold for different values of $\mu_{l}$.  When $\mu_{l}>|M-2J|$ the
oscillations are more pronounced due to the enhanced finite overlap of the initial perturbation with
bulk states. Otherwise, the conclusions stay as above.

Finally, we have also studied the decay of a local perturbation at a bulk site, as a function of the gap
size.  The resulting disturbance spreads out in all directions and there is no notion of chirality
in the propagation of the excitation.  We also find no correlation with the gap.  We attribute this
to the fact that a spatially localized bulk perturbation is strongly delocalized in momentum space
and thus it is insensitive to the size of the gap.

%%%%%%%%%%%% FIGURE %%%%%%%%%%%%%%%%% FIGURE  %%%%%%%%%%%%%%%%%% FIGURE %%%%%%%%  
\begin{figure}
   \centering
      \includegraphics[width=0.5\columnwidth]{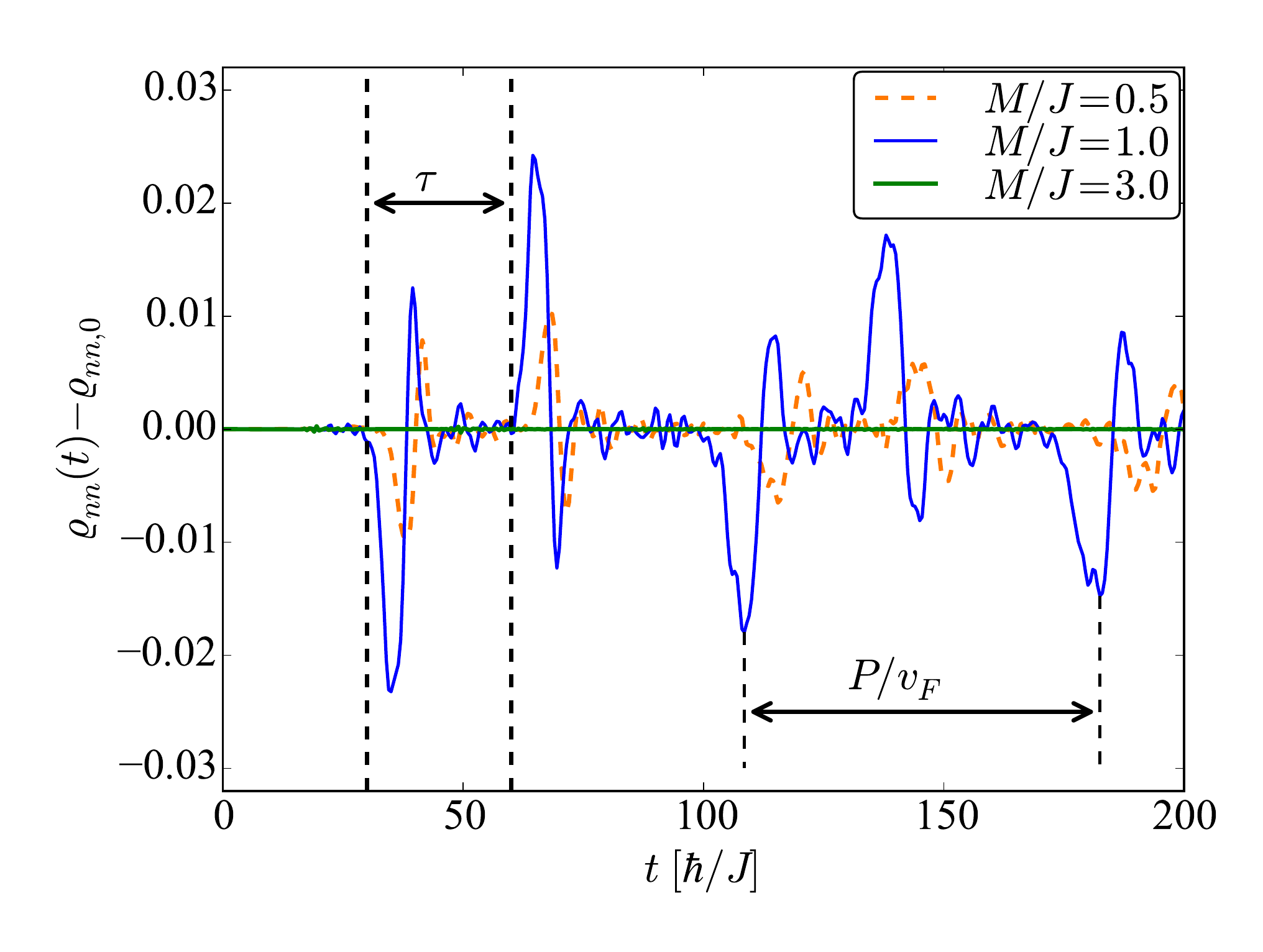} 
   \caption{(color online) The density difference $\varrho_{nn}(t)-\varrho_{nn,0}$ at the site
     labeled $n$ [see Fig.~\ref{fig:edgepertpulse1}] is plotted against time for the case of the
     pulse quench with $t_1=0$ and $t_2=30\hbar/J$ for parameters 
     $M/J=3$ (featureless green solid line), $M/J=1$ (blue solid line) and $M/J=1/2$ (orange dashed
     line). 
   % Same density difference as in Fig.~\ref{fig:edgepertpulse1} calculated at the perturbed edge site for different values of gap sizes $2|M-2J|$, parameterized by $M/J=1$ (blue solid line), $M=J/2$ (green dashed line). 
     As in Fig.~\ref{fig:edgepert2} there is a slight period delay as the gap decreases due to the
     finite extent of the edge state. The intervals $\tau=t_{2}-t_{1}$  as well as $T_{\text{edge}}\simeq P/v_F$, where $P$ is the perimeter of the lattice and $v_F$ is the Fermi velocity of the edge state, are indicated by vertical
     dashed lines.}
   \label{fig:edgepertpulse2}
\end{figure}
%%%%%%%%%%%% FIGURE %%%%%%%%%%%%%%%%% FIGURE  %%%%%%%%%%%%%%%%%% FIGURE %%%%%%%%  

\subsection{Local density pulse}

We now study a density perturbation pulse [see Fig.~\ref{fig:quenches}(b)] of width $\tau$,
\begin{equation}
H(t) = H_{0} + H_{1}^{l}\left[\theta(t-t_{1})-\theta(t-t_{2})\right] ,
\end{equation} 
with $H_{0}=H_{\mathrm{CI}}$ and  $\tau=t_{2}-t_{1}$.  We choose $t_{1}=0$ and $t_{2}=\tau$ without loss of generality.

We focus on a local edge perturbation (at a site $l\in$ edge) and look at the time evolution as a
function of $\tau$. Since the relevant energy scale is $J$ we expect two different regimes depending
on whether $\tau<\hbar/J$ or $\tau>\hbar/J$.  Indeed, in the former case ($\tau<\hbar/J$) the
perturbation is too localized in time to react separately to both the turning on of the perturbation
at $t_{1}$ and releasing at $t_{2}$.  Accordingly, the system generates a single signal traveling
along the edge of the system, which is allowed to conduct chirally due to the non-trivial topology
of the Hamiltonian. Such a disturbance evolves in time qualitatively as was shown in
Fig.~\ref{fig:edgepert} and thus is not shown here.

On the other hand, when $\tau>\hbar/J$ the system can react to both the pressing at $t_{1}$ and the
release at $t_{2}$. In this case, two pulses are generated, one at time $t_{1}$ when the system is
`pressed' (perturbation is turned on) and one at $t_{2}$ when the local density perturbation is
released [see Fig. \ref{fig:edgepertpulse1}].

The Fermi velocities of both pulses are set by $v_{F}=J$ since the arguments presented in the previous section still apply.
Similarly, by changing $M$ [see Fig.~\ref{fig:edgepertpulse2}] a slight delay in the wave packet center propagation, analogous to that observed in Fig.~\ref{fig:edgepert2},
is evident.

\section{\label{sec:trap}Chiral `edge' dynamics in power-law traps}

In this section, we consider fermions in a Chern-band lattice in the presence of a power-law trap 
\begin{subequations}
\begin{eqnarray}
H_{0}&=&H_{\mathrm{CI}} +H_{\mathrm{trap}},\\
H_{\mathrm{trap}} &=& J \sum_{i} \left(\dfrac{r_{i}}{r_{0}}\right)^{\gamma}c^{\dagger}_{i}c_{i},
\end{eqnarray}
\end{subequations}
where $r_{i}$ is the distance of site $i$ from the center of the trap, which in our simulations will
coincide with the center of the lattice.  The purpose of considering a trapped Chern lattice is to
make contact to their possible cold-atom realizations.

Cold atom experiments are generally performed in the presence of harmonic traps, i.e., $\gamma=2$
traps.  In such a trap, it may be difficult to distinguish between edge and bulk regions.  As a
result there is considerable interest in power-law traps with larger exponents, i.e., large
$\gamma$, because for large $\gamma$ a power-law trap can resemble a box trap with sharp boundaries,
and hence can be expected to have better-defined edge states spatially separated from the
bulk~\cite{GDD13,DG13,GSG13,MSH05}. For example, Refs.\ \onlinecite{GDD13} and \onlinecite{DG13}
propose non-equilibrium protocols involving hard boundaries for experimentally determining Chern
numbers, using the idea that approximate hard-wall potentials can be designed using power-law
potentials with large $\gamma$.

Motivated by this interest in large $\gamma$, in this section we present equilibrium spatial structures of
fermions loaded in a $\gamma=50$ trap, and based on this knowledge we explore chiral responses of
the `edges' of such fermionic clouds to local quenches.

%%%%%%%%%%%% FIGURE %%%%%%%%%%%%%%%%% FIGURE  %%%%%%%%%%%%%%%%%% FIGURE %%%%%%%%  
\begin{figure}
   \centering
      \includegraphics[width=0.5\columnwidth]{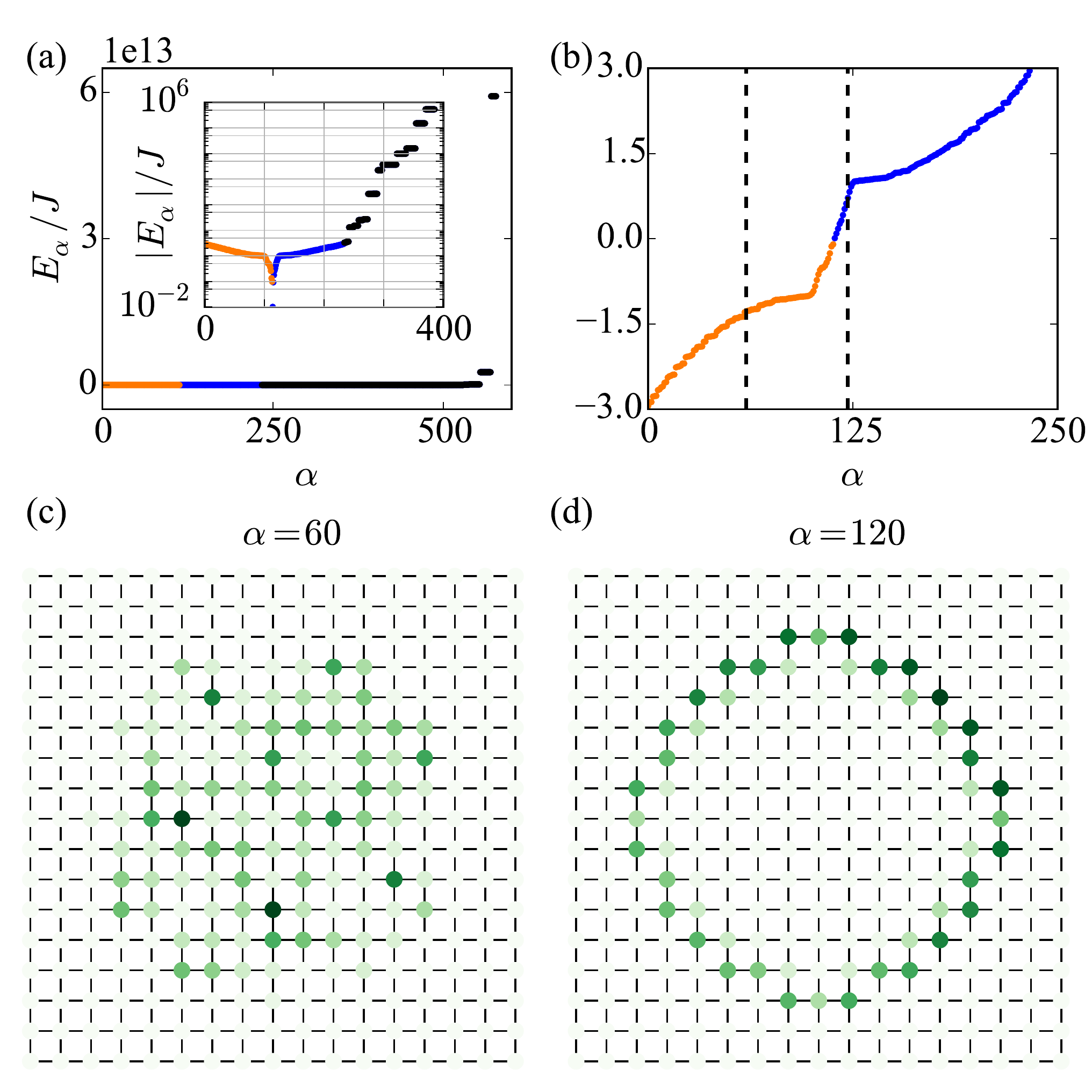} % requires the graphicx package
      \caption{(color online) Single-particle energy spectrum and eigenstates in the presence of a
        trap, $r_0=6a$ and $\gamma=50$, for $M/J=1$ and $L_x=L_{y}=17$.  (a) Energy spectrum.
        Inset shows absolute values on a logarithmic scale. (b) Magnification of low energy sector.
        The eigenvalues in this sector resemble that of a Chern insulator with open boundary
        conditions [compare with Fig.~\ref{fig:bands}(a)].  (c,d) Probability distributions for the
        $\alpha=60$ and $\alpha=120$ eigenstates (dashed lines in (b)), corresponding to a `bulk
        state' and an `edge state'.  }
   \label{fig:trapfig}
\end{figure}
%%%%%%%%%%%% FIGURE %%%%%%%%%%%%%%%%% FIGURE  %%%%%%%%%%%%%%%%%% FIGURE %%%%%%%%  

The effect of a large-$\gamma$ trap is to separate energetically single-particle eigenstates which
are spatially `inside' the trap (within distance $r_0$ from the trap center) from those `outside'
the trap.  A low-energy sector thus emerges which closely resembles a uniform system of radius
$r_0$.  By `half-filling' this region, i.e., by having $N\sim\pi{r^2_0/a^2}$ fermions in the entire
system, one can then mimic a half-filled region with a reasonably well-defined edge.  Such an `edge'
also shows chiral dynamics as in the open-boundary case without a trap treated in the previous
section. 

In Fig.~\ref{fig:trapfig}, we use a square lattice with sides larger than $2r_0$, and $\gamma=50$.
The top panels show the single-particle eigenspectrum.  The higher energy sectors contain
eigenstates whose weights are spatially concentrated in regions $r>r_0$.  The zoom onto the
low-energy sector [see top right panel in Fig.~\ref{fig:trapfig}] resembles the spectrum of a Chern lattice without a trap and with
open boundary conditions. Indeed, eigenstates with $-1<E/J<1$ have an edge-like distribution localized around 
$r\sim r_{0}$, exemplified by the lower right panel in Fig.~\ref{fig:trapfig}.
On the other hand, eigenstates with $1<|E/J|<3$ are extended over $r<r_0$, resembling the 
bulk behavior of a Chern insulator [see bottom left panel in Fig.~\ref{fig:trapfig}].

Note that the midgap 'edge' states with $-1<E/J<1$ are not as sharply defined as in the free Chern insulator case, as can be seen by
comparing the lower panels in Fig. \ref{fig:trapfig} with Fig. \ref{fig:bands} (a).  This is because,
even at such large $\gamma$, the lack of a hard wall induces mixing between eigenstates with edge
and bulk character.  Nevertheless, the edge modes are well-defined enough to display chirality in
real-time dynamics, as we next show.

\begin{figure}
   \centering
      \includegraphics[width=0.5\columnwidth]{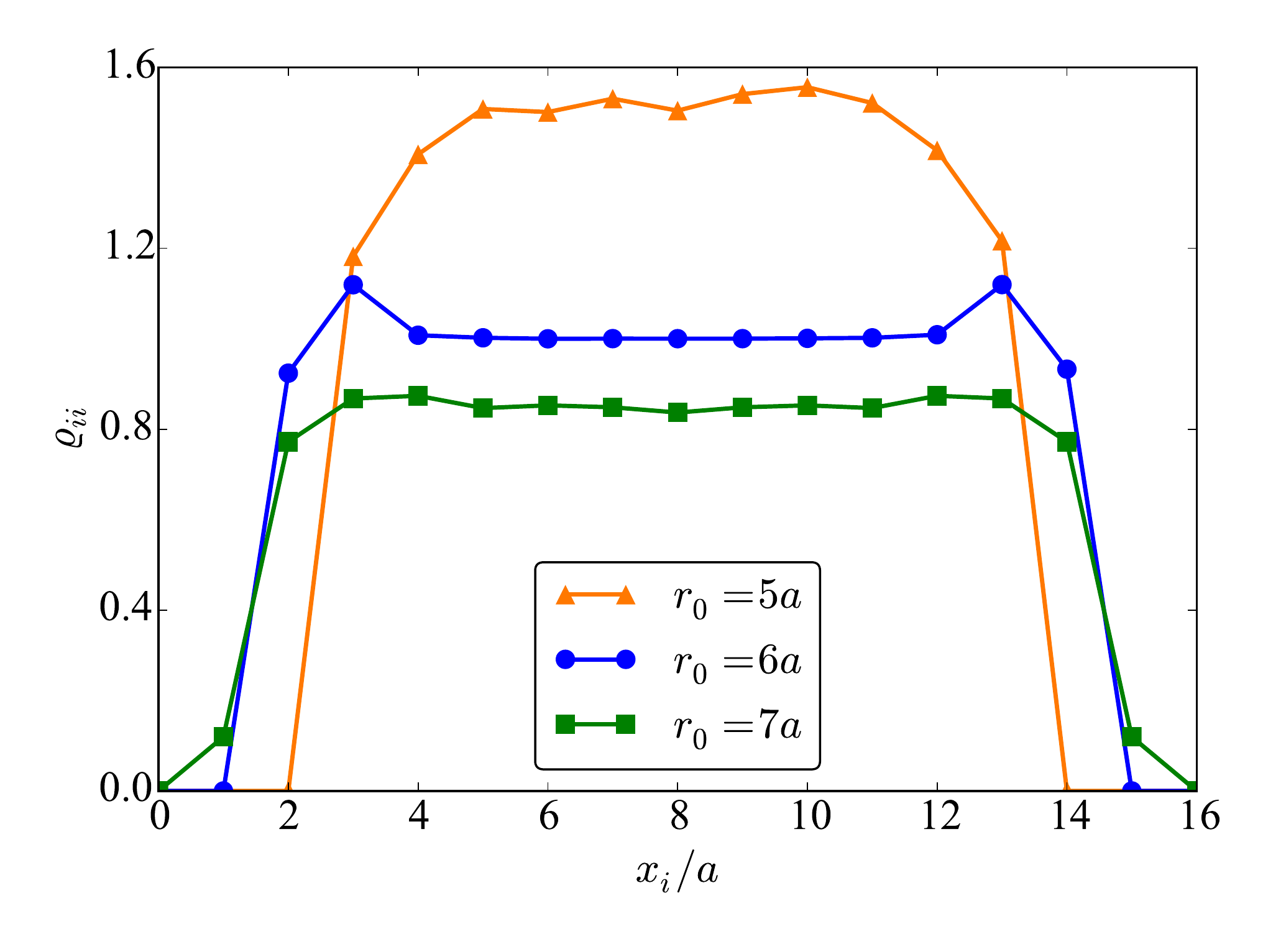}
      \caption{(color online) The density along the $x$ direction for a fixed cut $y=8a$ in real
        space for different values of $r_0$ for the ground state with $N=120\sim\pi6^2$ fermions,
        $\gamma=50$, $M/J=1$ and $L_x=L_{y}=17$.  For this $N$, the edge states are only
        visible for $r_0=6a$ as bumps in the density at the edge of the trapped cloud.}
   \label{fig:r0effect1}
\end{figure}

To access the chiral character of these effective edge states, it is necessary to mimic the situation of
half-filling.  This is achieved by having the fermion number to be close to $\pi{r^2_0/a}^2$.  In
Fig.~\ref{fig:r0effect1} we fix the fermion number to be $N=120\sim\pi{6}^2\simeq113$, and compare the
density profiles along the horizontal ($x-$) direction for traps with $r_0=5a$, $r_0=6a$, and
$r_0=7a$.  The figure shows that this particular filling provides enough fermions to occupy the
effective edge states when $r_0=6a$. The edge occupancy is only visible in the $r_0=6a$ case through the bumps in the density profile peaked at
$x=3a$ and $x=13a$; the other curves lack this feature.  This exemplifies the fact that, depending on the trap
shape (set by $r_{0}$), there is an optimal particle number for accessing the chiral edge states of
the system.

%%%%%%%%%%%% FIGURE %%%%%%%%%%%%%%%%% FIGURE  %%%%%%%%%%%%%%%%%% FIGURE %%%%%%%%  
%%%%%%%%%%%% FIGURE %%%%%%%%%%%%%%%%% FIGURE  %%%%%%%%%%%%%%%%%% FIGURE %%%%%%%%  

\begin{figure*}
   \centering
      \includegraphics[width=0.99\linewidth]{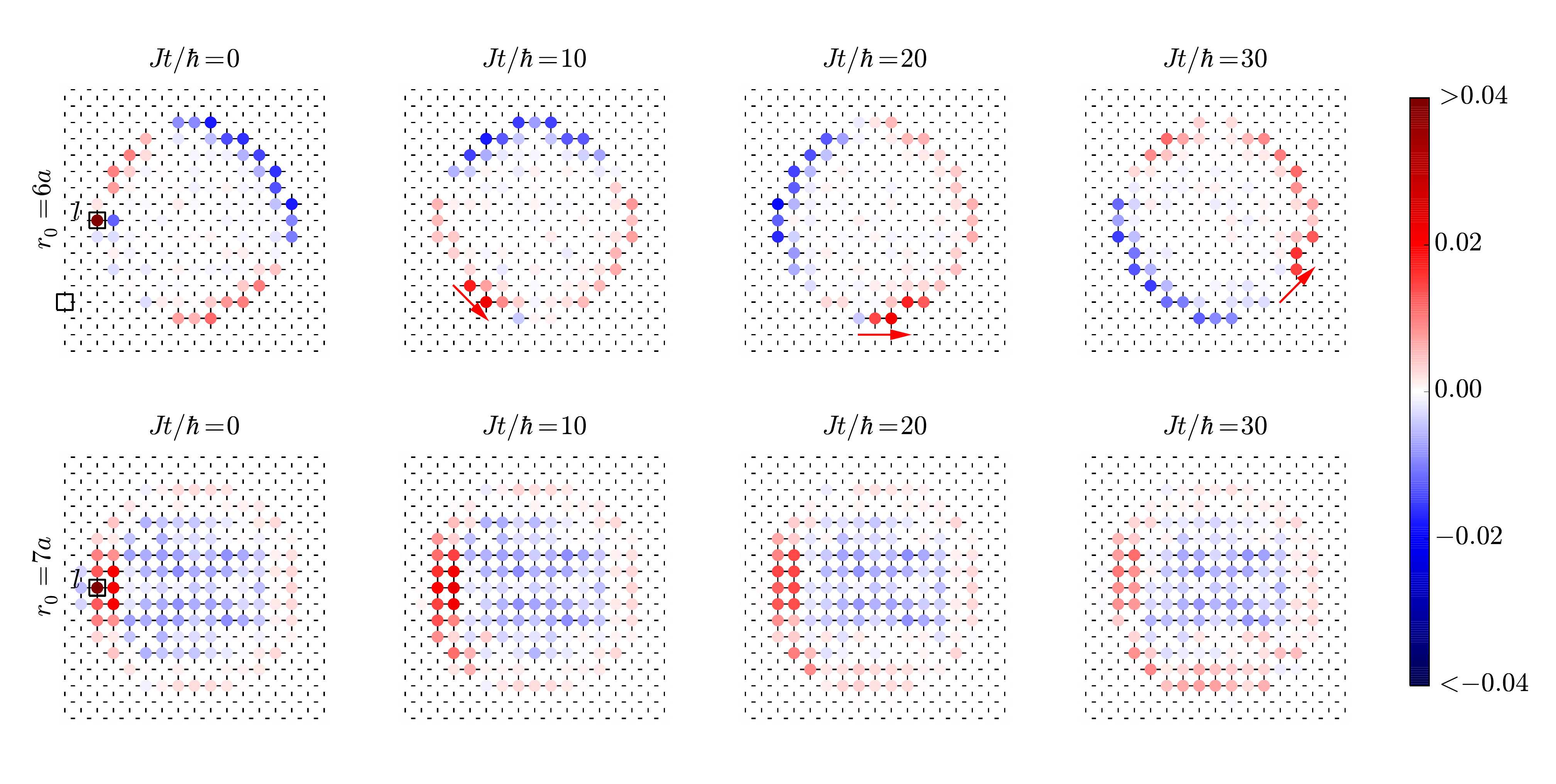} 
      \caption{(color online) Four snapshots of the density perturbation
        $\varrho_{ii}(t)-\varrho_{ii,0}$ after a quench of the form of Fig.~\ref{fig:quenches}(a),
        for the Hamiltonian $H_{\mathrm{CI}}$ with $|M|< 2J$ in the presence of a $\gamma=50$ trap.
        Trap radius is $r_{0}=6a$ (upper panels) or $r_{0}=7a$ (lower panels).  The simulation
        lattice is $17a\times17a$; the quench strength is $\mu_{n}=-J/4$. The particle number
        $N=120\approx \pi6^2$ is chosen such that the last filled eigenvalue for $r_{0}=6$ is that
        of Fig. \ref{fig:trapfig}(d).  Here $\varrho_{ii,0}$ corresponds to the site densities in
        the ground state of $H_0=H_{\mathrm{CI}}+H_{\mathrm{trap}}$ filled with $N=120$ fermions. Please see supplementary material for a video of the time evolution.}
   \label{fig:r0effect2}
\end{figure*}

In order to highlight the importance of the filling fraction with respect to the trap parameters, in
Fig.~\ref{fig:r0effect2}, we show the real-time evolution to a perturbation at a fixed site $l$
using the non-equilibrium protocol of Fig.~\ref{fig:quenches}(a) for $r_{0}=6a$ and $r_{0}=7a$.  The
particle number is fixed for both cases to be $N=120 \sim \pi{6}^2\approx113$, the same value as in
Fig.~\ref{fig:trapfig}.  Therefore the edge states are only populated for $r_0=6a$ but not for
$r_0=7a$.  For $r_0=6a$ (upper panels of Fig.~\ref{fig:r0effect2}), the perturbation follows the
profile of the trap in a chiral fashion, a situation similar to a half-filled open-boundary system.
For $r_0=7a$, the density disturbance does not propagate along the new edge, but dissipates into the
bulk.

Note that there is some flexibility in varying $N$ for a fixed $r_0$ (and vice versa) while
  still obtaining a visible edge state.  As long as the state at the Fermi energy is one of the edge
  states, the physics of edge states is accessible.  The number of edge states is approximately
  $2\pi r_0$, the circumference of the interior region.  The lattice geometry will affect the exact
  number, of course, but for $r_0\gg1$, this is a reasonable estimate.  For smaller $r_0$, the
  spectrum in Figure 7 exemplifies the situation: in this case ($r_0=6a$), the edge states run from eigenstate
  $\alpha\sim95$ to eigenstate $\alpha\sim131$.  The physics of edge states is visible as long as
  $N$ is in this range.

For the $r_0=6a$ case, as the gap is decreased by changing the ratio $M/J$ 
appropriately, the effective edge states become less localized. Fixing the rest of parameters 
we observe a retardation effect of the wave packet, consistent with that discussed in Sec. \ref{sec:half-filling} for the evolution in 
the absence of $H_{\mathrm{trap}}$.

For completeness we have also investigated the protocol in Fig.~1 (b) under the effect of the trap
and found similar results to the scenario without a trap as long as the system is close to the
optimal filling discussed above.  This protocol generates two pulses confined to the boundary of the
trap with similar properties as those discussed in previous sections.  Finally we expect that the
effect of softer traps, i.e. smaller values of $\gamma$, addressed for instance in
Refs.~\onlinecite{SGD10,GDD13}, will result in an overall broadening of the edge states and greater
mixing between bulk and edge.  To what extent the chirality in real-time dynamics is visible for
smaller $\gamma$ remains an open question.

%%%%%%%%%%%% FIGURE %%%%%%%%%%%%%%%%% FIGURE  %%%%%%%%%%%%%%%%%% FIGURE %%%%%%%%
%%%%%%%%%%%% FIGURE %%%%%%%%%%%%%%%%% FIGURE  %%%%%%%%%%%%%%%%%% FIGURE %%%%%%%%  

\section{\label{sec:conc}Discussion and Conclusions}

In this work we have investigated the effect of local edge quenches on the dynamics of Chern
bands. We have addressed two types of local density quench protocols to probe directly the chirality
of the edge states, both with and without confining traps.  When the bands are topologically
non-trivial, these quenches generate localized edge excitations that propagate chirally along the
sample with a group velocity that decreases as the gap controlling the spatial
delocalization of the edge modes decreases. In addition, we have shown that, in the presence of a sharp trap
with a characteristic length scale $r_{0}$, the chiral dynamics is only observable if the filling is
such that the effective trap confinement region is half-filled.  These results emphasize the
importance of the effective filling within the trap to observe edge dynamics, even when the trap
potential is quite sharp.

We have concentrated on large-exponent power-law traps (large $\gamma$).  While this is not the
common situation in current cold-atom experiments, interest in producing and utilizing such traps is
high, particularly in the context of topological matter~\cite{GDD13,DG13,GSG13,MSH05}. We have shown
data for $\gamma=50$, but it is expected that traps with exponents down to $\gamma\approx4$ will
have similar properties~\cite{KGF14a}.  By analyzing the spectrum, we have shown that the `inside'
and `outside' of the trap are energetically separated.  The `inside' region is found to cover
${\pi}r_0^2$ lattice sites.  This observation has allowed us to specify the particle number which
mimics the physics of half filling in the `inside' region.  We have shown that the chiral dynamics
is not visible when the number of fermions is very different from this optimal, because the effective
filling then differs from half-filling.  

It is plausible that the protocols considered in this work can be realized in cold atomic set-ups.
Addressing and imaging these systems with single-site resolution have become available in cold-atom
laboratories during the past few years~\cite{BPT10, SWE10, WES11}.  Thus, performing local quenches
on the recently realized Chern-band lattices~\cite{DG13,AAL13,ALS14} and following the ensuing site
density dynamics in real time should be technologically feasible.

Our findings might also be relevant in solid state set-ups.  For instance, the Chern insulator state
has been recently realized by magnetic doping a thin-film structure of a three-dimensional
topological insulator~\cite{CZJ13}.  In this context, the protocol in Fig.~\ref{fig:quenches}(a),
theoretically can be thought of as an STM tip perturbing the system locally to then release the
density. The typical time scales governing such dynamics in this case are several orders of
magnitude faster than in cold atomic experiments, rendering such a proposal practically unviable
experimentally. However, fast, out-of-equilibrium photo excitation of electrons has been measured
with recent pump-probe techniques~\cite{WSJ13}.   In this experiment, the local dynamics occurring
within typical electron time-scales of femto-seconds were measured at the surface of a
three-dimensional topological insulator.  In light of these results we can reinterpret the protocol
in Fig.~\ref{fig:quenches}(b) in a first approximation as a laser pulse of duration $\tau$ that
probes the electrons locally in space and time.  It is therefore not unrealistic to think
that implementing such a protocol might be possible by probing the Chern insulator state of
Ref.~\onlinecite{CZJ13}, especially considering that they are grown from essentially the same family
of materials that were pump-probed~\cite{WSJ13}.

\section*{Acknowledgments}

The authors thank P.~McClarty, R.~Moessner, and P.~Riberio for useful discussions. 

\appendix

\section{\label{app:evol}Density matrix time evolution}

Time evolution in this work has been performed by evolving the single-particle density matrix, i.e.,
the matrix of two-point correlators.  The method is standard for systems described by quadratic
fermions and is widely used in the non-equilibrium literature, so we outline the procedure here only
briefly.  The initial equilibrium density operator is
$\varrho=\sum_{l}\left|\psi_{l}\right\rangle\left\langle\psi_{l}\right|$, where the sum is over
occupied single particle eigenstates $\left|\psi_{l}\right\rangle$ of the initial Hamiltonian.  

The time evolution of such an operator is given by
\begin{equation}  \label{eq:densmatr_evolution}
\varrho(t)=e^{-iHt}\varrho e^{iHt}.
\end{equation}
where $H$ is the single-particle Hamiltonian under which the evolution occurs.  As single-particle
operators, the density and Hamiltonian operators can each be represented as
$2\times{L_x}\times{L_y}$ matrices in our case.  For the lattice sizes we have used, the matrix
dimensions are $<10^{3}$; the matrix operations of Eq,.\ \eqref{eq:densmatr_evolution} are thus
numerically inexpensive.  The $\varrho(t)$ matrices at each time can thus be performed either by
rotating to the basis where $H$ is diagonal, or even by explicit computation of the matrix
exponentials.  For the pulse case, the Hamiltonian changes at the end of the pulse; we perform
evolution with Hamiltonian $H_0+H_1$ for time $\tau$ and then switch to evolving with the
Hamiltonian $H_0$.

Physically, the diagonal elements of $\varrho_{ii}(t)$ represent the evolution of the density at
site $i$ with the constraint that $\mathrm{Tr}[\varrho(t)]=N$ where $N$ is the number of particles.
The off diagonal elements $\varrho_{ij}(t)$ with $i\neq j$ are equal-time correlation functions at
time $t$ between sites $\left\lbrace i,j\right\rbrace$.

\end{document}